\documentclass[useAMS,usenatbib,usegraphicx]{mn2e}
\usepackage{todonotes}
\title[]{HIP\,3678: a hierarchical triple stellar system in the centre of the planetary nebula NGC\,246\thanks{Based on observations obtained at the Paranal Observatory in ESO programmes 074.D-0315(A), and 079.D-0546(A), as well as with the NASA/ESA \textit{Hubble Space Telescope} in programme 5106.}} \author[C. Adam and M. Mugrauer]{C. Adam$^{1}$\thanks{E-mail: christian.adam.1@uni-jena.de} and M. Mugrauer$^{1}$\\ $^{1}$Astrophysical Institute and University Observatory, Schillerg\"{a}{\ss}chen 2, D-07745 Jena, Germany}
\begin{document}

\date{}

\pagerange{\pageref{firstpage}--\pageref{lastpage}} \pubyear{2014}

\maketitle

\label{firstpage}

\begin{abstract}

We report the detection of a new low-mass stellar companion to the white dwarf HIP\,3678\,A, the central star of the planetary nebula NGC\,246. The newly found companion is located about 1\,arcsec (at projected separation of about 500\,au) north-east of HIP\,3678\,A, and shares a common proper motion with the white dwarf and its known comoving companion HIP\,3678\,B. The hypothesis that the newly detected companion is a non-moving background object can be rejected on a significance level of more than 8\,$\sigma$, by combining astrometric measurements from the literature with follow-up astrometry, obtained with Wild Field Planetary Camera 2/\textit{Hubble Space Telescope} and NACO/Very Large Telescope. From our deep NACO imaging data, we can rule out additional stellar companions of the white dwarf with projected separations between 130 up to 5500\,au. In the deepest high-contrast NACO observation, we achieve a detection limit in the \textit{Ks} band of about 20\,mag, which allows the detection of brown dwarf companions with masses down to 36$\,\rmn{M_{jup}}$ at an assumed age of the system of 260\,Myr. To approximate the masses of the companions HIP\,3678\,B and C, we use the evolutionary \citeauthor{baraffe1998} models and obtain about 0.85\,M$_{\sun}$ for HIP\,3678\,B and about 0.1\,M$_{\sun}$ for HIP\,3678\,C. According to the derived absolute photometry, HIP\,3678\,B should be a early to mid-K dwarf (K2--K5), while HIP\,3678\,C should be a mid-M dwarf with a spectral type in the range between M5 and M6.

\end{abstract}


\section{Introduction}

NGC\,246, also known as the `Skull Nebula' is a planetary nebula in the constellation Cetus, which was first observed by W.~Herschel in 1785. The nebula exhibits a slightly elliptical morphology with a diameter of about 224\,arcsec in average \citep{cahn1992}, a temperature of about 20000\,K, and shows Ne\,{\sc v} and O\,{\sc vi} emission lines in its far-ultraviolet spectrum, as measured with the \textit{Far-Ultraviolet Spectroscopic Explorer} (\textit{FUSE}), indicating photoionization from the intense UV-radiation of its central star \citep{hoogerwerf2007}. The nebula is classified as an old WZO3 stage interacting planetary nebula with an dynamical age of about 6600\,yr \citep[and references therein]{ali2012}, which expands with a velocity of about 39.5\,km/s, as derived by \citet{weinberger1989}.

The central star of the planetary nebula is HIP\,3678\,A (alias WD\,0044-121), whose proper and parallactic motion was measured by the astrometry satellite \textit{Hipparcos} \citep[$\mu_{\alpha}\cos(\delta)=-23.85\pm3.42$\,mas/yr and $\mu_{\delta} = -4.89\pm1.82$\,mas/yr, $\pi = 2.12\pm3.01$\,mas, as newly determined by][]{vanleeuwen2007}. HIP\,3678\,A is a very hot O\,{\sc vi} sequence \citep{smith1969}, or PG\,1159-35(lg\,E) \citep{werner2006} star with an effective temperature of about 150000\,K, a mass of 0.84\,$\rmn{M}_{\sun}$, and a surface gravity of log(g) $\sim5.7$\,(cgs) \citep{koesterke1998}. Whereas the nebula was not detected in X-ray, its central star has \textit{ROSAT} and \textit{Chandra} detections and shows a non-local thermodynamical equilibrium model consistent PG\,1159-type spectrum \citep{green1986}.

HIP\,3678\,A has a known common proper motion companion, which is located about 3.8\,arcsec northeast of its primary star, first noted by \citet{minkowski1965} and later then confirmed by \citet{cudworth1973}. By fitting photometric measurements of the comoving companion to the zero age main sequence, \citet{bond1999} derived a distance of $495^{+145}_{-100}\,$pc for the HIP\,3678 system, while \cite{terzian1997} derived a distance of $570^{+155}_{-150}$\,pc, using the parallax expansion method, based on imaging data of the planetary nebula, obtained by \cite{liller1966}. Additional distances of the planetary nebula were obtained \citep[$480 \pm 96$\,pc, and $472 \pm 670$\,pc, by][and McDonald et al. 2012, respectively\nocite{mcdonald2012}]{stanghellini2010}, which well agree with the previous distance estimates. In our research we adopt here a distance of $504 \pm 178$\,pc, which is the mean and the $1\,\sigma$-error of the distance values given in the literature, as listed above.

The planetary nebula NGC\,246 with its central stellar system is shown in the right-hand panel of Fig.\,\ref{fig:imags}. This pattern is a colour-composite image, composed of images from the Digitized all Sky Survey (DSS), which were taken through different optical filters at observing epochs between 1954 and 1994.
Assuming a distance of 504\,pc, NGC\,246 exhibits an averaged projected diameter of about 110000\,au. The elliptical shape of the planetary nebula is clearly visible in the DSS image with its semimajor axis aligned in the east to west direction, consistent with the motion of its central stellar system. The (leading) western shell of NGC\,246 appears clearly brighter than the eastern (trailing) shell of the nebula, possibly induced by compressional heating due to the motion of the planetary nebula through space and its interaction with the interstellar medium.

In this paper, we present astro- and photometric measurements of the stellar system HIP\,3678 in the centre of NGC\,246, which were obtained and analysed in the course of our high-contrast imaging survey, to study the multiplicity of B stars in the near-infrared, using imaging data taken with the adaptive optics imager NACO \citep[][and Rousset et al. 2003]{lenzen2003} \nocite{rousset2003} at the Very Large Telescope (VLT), operated by the European Southern Observatory (ESO) in Chile.

The order of this paper is as follows. In Section~\ref{sec:observations}, we describe all details of the individual observations and the applied data-reduction procedures, and present the astrometric calibration of the used instruments. In the Section~\ref{sec:astrometry}, we show all astrometric measurements, as well as the proper motion analysis of the detected companions in the HIP\,3678 system. In Section~\ref{sec:phot}, we present the results of our photometric analysis, as well as the mass and age estimation of the detected companions. Finally, in the last section we summarize and discuss all results reported in this work.

\section{Observations and Data Reduction}
\label{sec:observations}

The near-infrared data of HIP\,3678, which are presented here, were taken with NACO in two observing epochs in 2004 and 2007. In 2004, HIP3678 was imaged with NACO's S13 optics in the \textit{J}, \textit{H}, and \textit{Ks} band, while in 2007 observations were taken only in the \textit{Ks} band using NACO's S27 optics. In both observing epochs, the jitter technique was applied to effectively cancel out the bright background of the sky in the near-infrared. Several short integrations (DIT) were taken per jitter position and several of these integrations (NDIT) were then averaged to one image. The target was then observed at several (NINT) different randomly chosen jitter positions located within a jitter-width of 6\,arcsec with the S13, and 5\,arcsec with the S27 optics.

\begin{figure*}
\resizebox{\hsize}{!}{\includegraphics{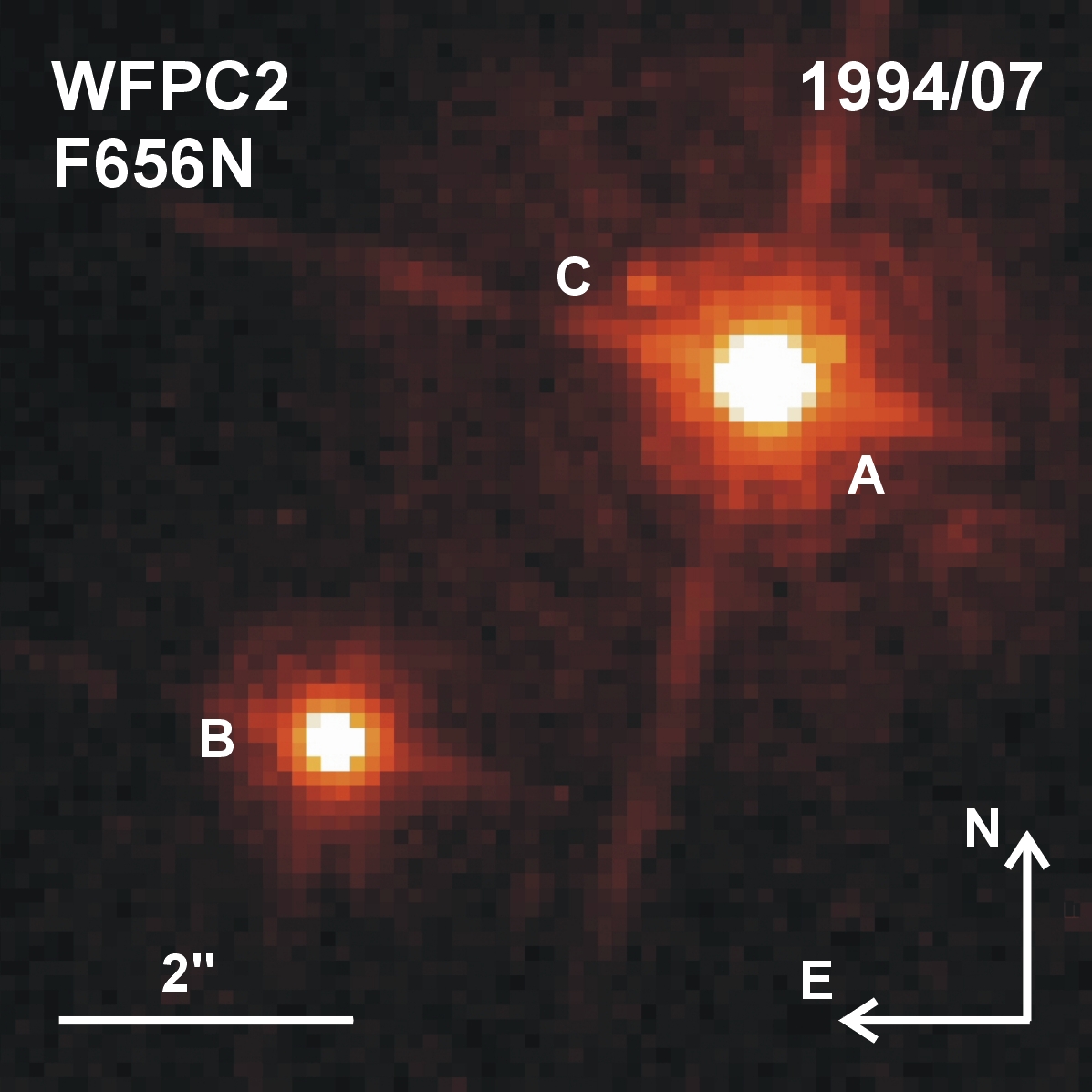} \includegraphics{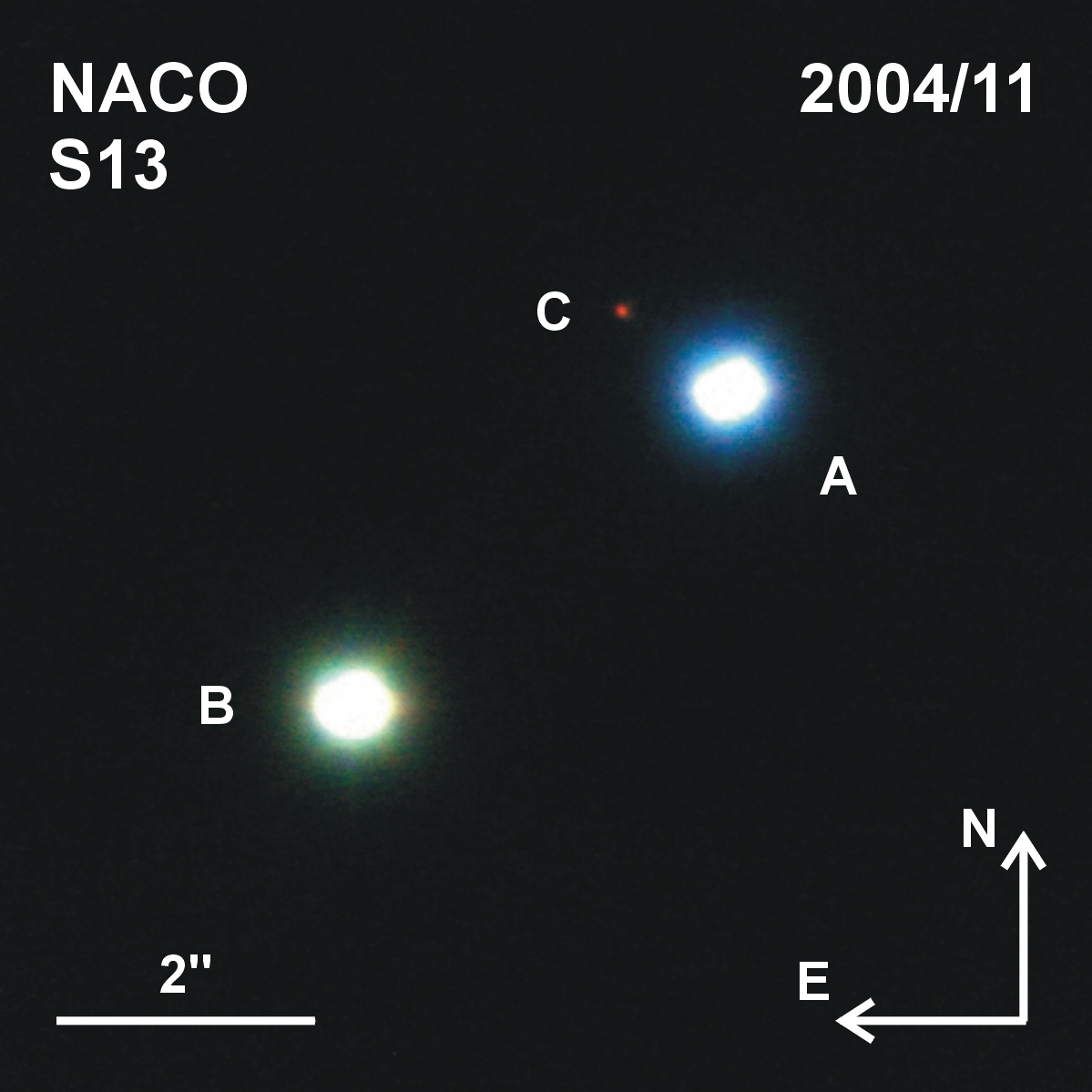} \includegraphics{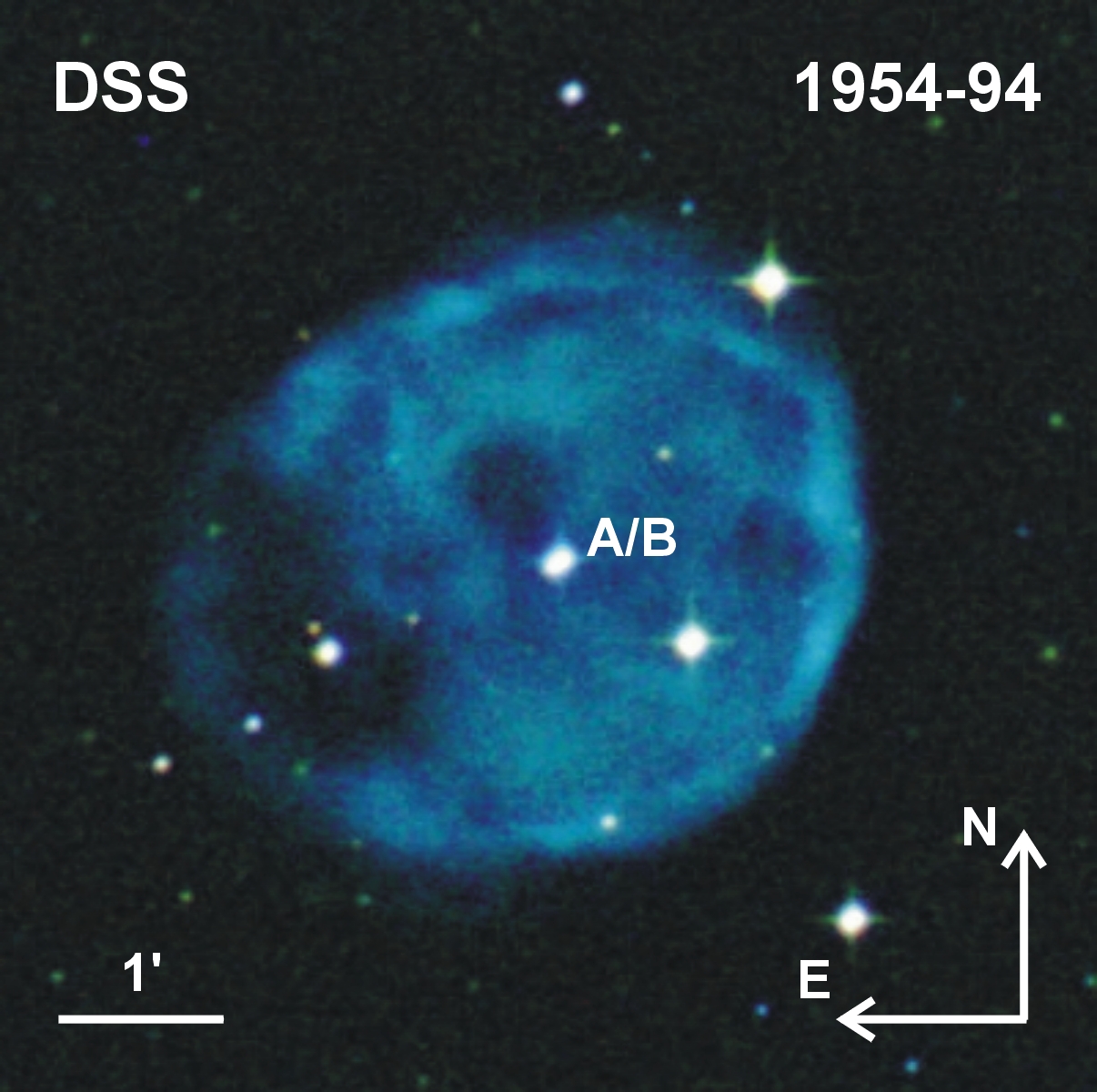}}

\caption{This pattern shows the fully reduced images of HIP\,3678, taken with WFPC2 (left) in the visual in 1994, the \textit{JHKs} band colour-composite image taken with NACO in 2004 (middle), as well as the colour-composite DSS-image (blue channel: POSSI.O-DSS2.706 (0.41\,$\mu$m) taken in 1954;
green channel: SERC ER DSS2.S681 (0.64\,$\mu$m) taken in 1989; red channel: SERC I DSS2.S681 0.807\,$\mu$m taken in 1994) of the planetary nebula NGC\,264 with the stellar system HIP\,3678 in its centre. The slightly elliptical shape of the planetary nebula is clearly visible in the colour-composite DSS image with its semi-major axis aligned in the east to west direction, induced by the motion of the system through space. The western leading edge of the nebular shell is brighter due to its interaction with the interstellar medium. In the WFPC2 and NACO images the individual components of the HIP\,3678 system in the centre of NGC\,246 are marked with letters.}\label{fig:imags}
\end{figure*}

In addition to the high-contrast near-infrared NACO observations, HIP\,3678 was also observed with the \textit{Hubble Space Telescope} (\textit{HST}) in 1994 in one observing epoch using the Wide Field Planetary Camera 2 (WFPC2). The fully reduced and calibrated image with a total detector integration time of 2000\,s, taken in the \textit{F656N} filter, was extracted from the \textit{HST} data-archive. Figure\,\ref{fig:imags} shows the NACO and \textit{HST} images of HIP\,3678 in the centre of NGC\,246.

For the data reduction of the NACO images, we use appropriate calibration data from the ESO data-archive, i.e. darks with the same exposure time, and flats taken in the same filter and in the same night as the science data. In order to reduce all imaging data, we use {\small ESOREX}\footnote{http://www.eso.org/sci/software/cpl/esorex}, which is part of ESO's Common Pipeline Library\footnote{http://www.eso.org/sci/software/cpl} (CPL). For each science frame a master-dark subtraction and flat-field correction is applied, and all images are averaged with {\small ESOREX}, using the provided shift+add procedure including measurement and subtraction of the bright background of the sky in the near-infrared.

All NACO images were astrometrically calibrated [determination of the pixel-scale (PS) and the position angle (DPA) of the NACO detector] using our astrometrical self-calibration technique, as described in detail by \citet{adam2013}. Thereby, the individual NACO science frames are used and shifts induced by the jitter technique are measured in the images, using all detected sources in the images. By comparing these shifts with the offsets of the telescope pointing, as given in the FITS-headers of the individual science frames, the pixel-scale and position angle of the detector can be determined even in the case that no astrometric standards are available for an observing epoch (which is the case here).

In contrast to the NACO observations, HIP\,3678 was observed with the \textit{HST} only at one telescope pointing, hence no astrometrical calibration of the WFPC2 detector for the given observing epoch is available. Therefore, we use here the pixel-scale and detector alignment, as given in world coordinate system (WCS) in the FITS-header of the \textit{HST} image, extracted from the \textit{HST} data-archive. Since no uncertainties for the astrometrical calibration are given, the resulting astrometric precision has to be considered as a lower limit.

The used instruments, filters, integration-times, number of taken images, as well as the derived astrometric calibration for all observing epochs, whose data are presented in this work, is summarized in the observation log in Tab.\,\ref{tab:obs}.

\begin{table*}

\caption{Observation log and astrometric calibration of all instruments and for all observing epochs, whose data are presented in this work. All NACO images are astrometrically calibrated using our self-calibration technique, while the pixel-scale (PS) and the detector position angle (DPA) of the \textit{HST} observation are taken from WCS of the \textit{HST} FITS-file.}
 \label{tab:obs}
 \begin{tabular}{lccccc}
	\hline
	Date           & Camera        & Filter & Exposure time & PS & DPA \\
                   &               &        & DIT [s] $\times$ NDIT $\times$ NINT & (mas pixel$^{-1}$) & ($\degr$)\\
\hline
	1994-07-27     & WFPC2/$HST$     & $F656N$  & 1000\,\,$\times$ \,\,\,1 $\times$ \,\,\,2\,                & $99.9$             & $0.0$\\
	2004-11-22     & NACO-S13/VLT  & $J$      & \,\,\,\,\,\,\,\,9 $\times$ \,\,\,8 $\times$ \,\,\,5     & $13.189 \pm 0.153$ & $+ 0.012 \pm 0.662$\\
	               & NACO-S13/VLT  & $H$      & \,\,\,3.6 $\times$ 20 $\times$ \,\,\,5                  & $13.259 \pm 0.118$ & $- 0.027 \pm 0.515$\\
	               & NACO-S13/VLT  & $Ks$     & \,\,\,6.1 $\times$ 20 $\times$ \,\,\,5                  & $13.182 \pm 0.117$ & $- 0.082 \pm 0.496$\\
	2007-06-14     & NACO-S27/VLT  & $Ks$     & \,\,\,\,40\, $\times$ \,\,\,5 $\times$ 15               & $27.285 \pm 0.190$ & $+ 0.153 \pm 0.403$\\
	\hline
 \end{tabular}

\end{table*}

\section{Astrometry}
\label{sec:astrometry}

Beside the known bright comoving companion HIP\,3678\,B, a further faint companion-candidate is detected in the available WFPC2 and NACO images of the central star of NGC\,246. The newly detected companion-candidate is located about 1\,arcsec north-east of HIP\,3678\,A. We measure the astrometric position of HIP\,3678\,A \& B, as well as that of the companion-candidate in all imaging epochs, using the {\small IDL}\verb"/starfinder" routine. With the astrometric calibration of each observing epoch, the angular separation and position angle of HIP\,3678\,B and of the companion-candidate relative to HIP\,3678\,A are derived, which are summarized in Tab.\,\ref{tab:obs}.

With the known proper and parallactic motion of HIP\,3678\,A, as well as the given epoch differences, we determine the expected angular separations and position angles of HIP\,3678\,B and of the newly detected faint companion-candidate for all observing epochs, assuming that these sources would be non-moving background objects, using the last observing epoch as reference.

\begin{figure*}
\Large{\underline{Proper motion diagram for HIP\,3678\,B}}
\vspace{0.4cm}

\resizebox{\hsize}{!}{\includegraphics{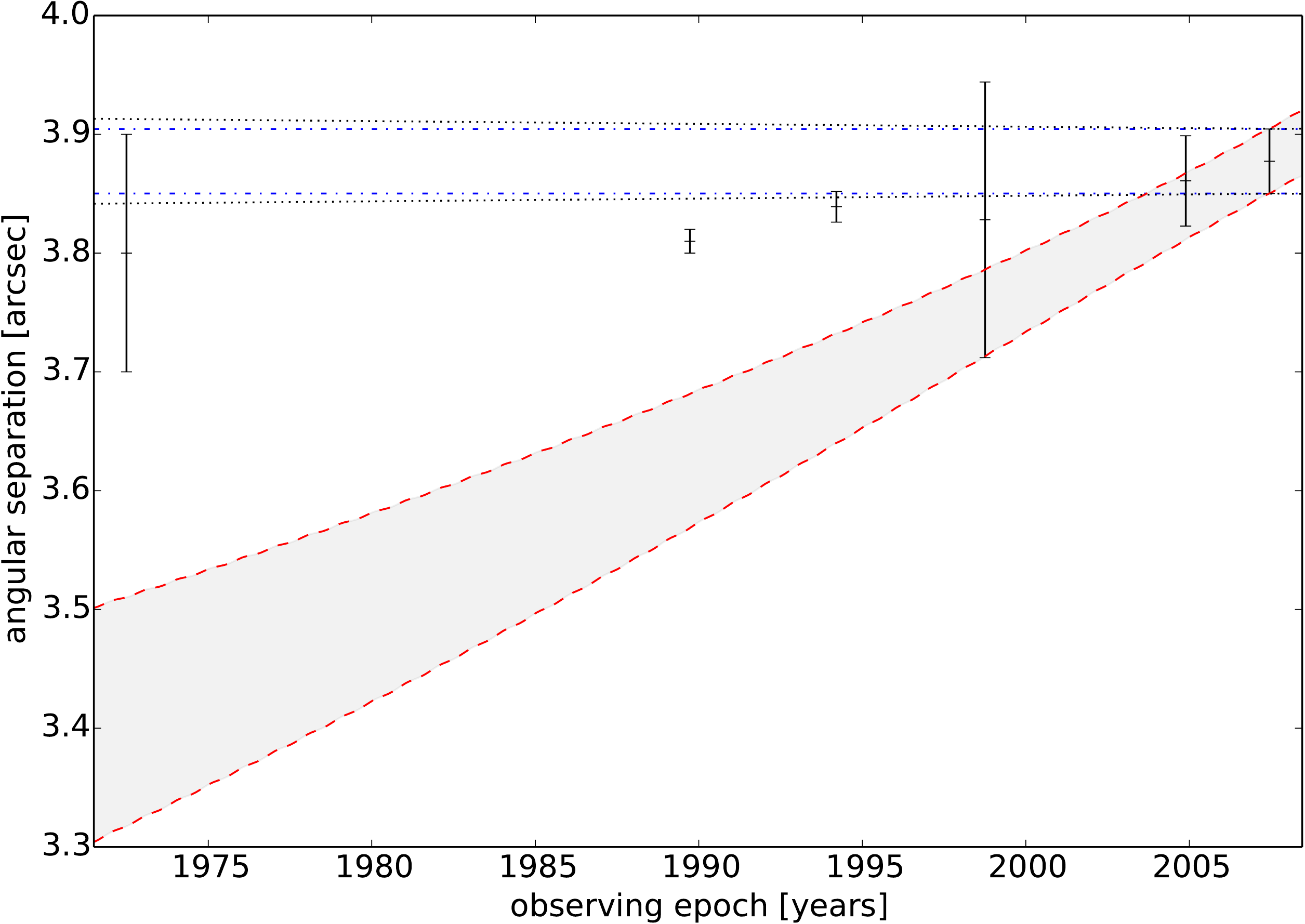}\hspace{10mm}\includegraphics{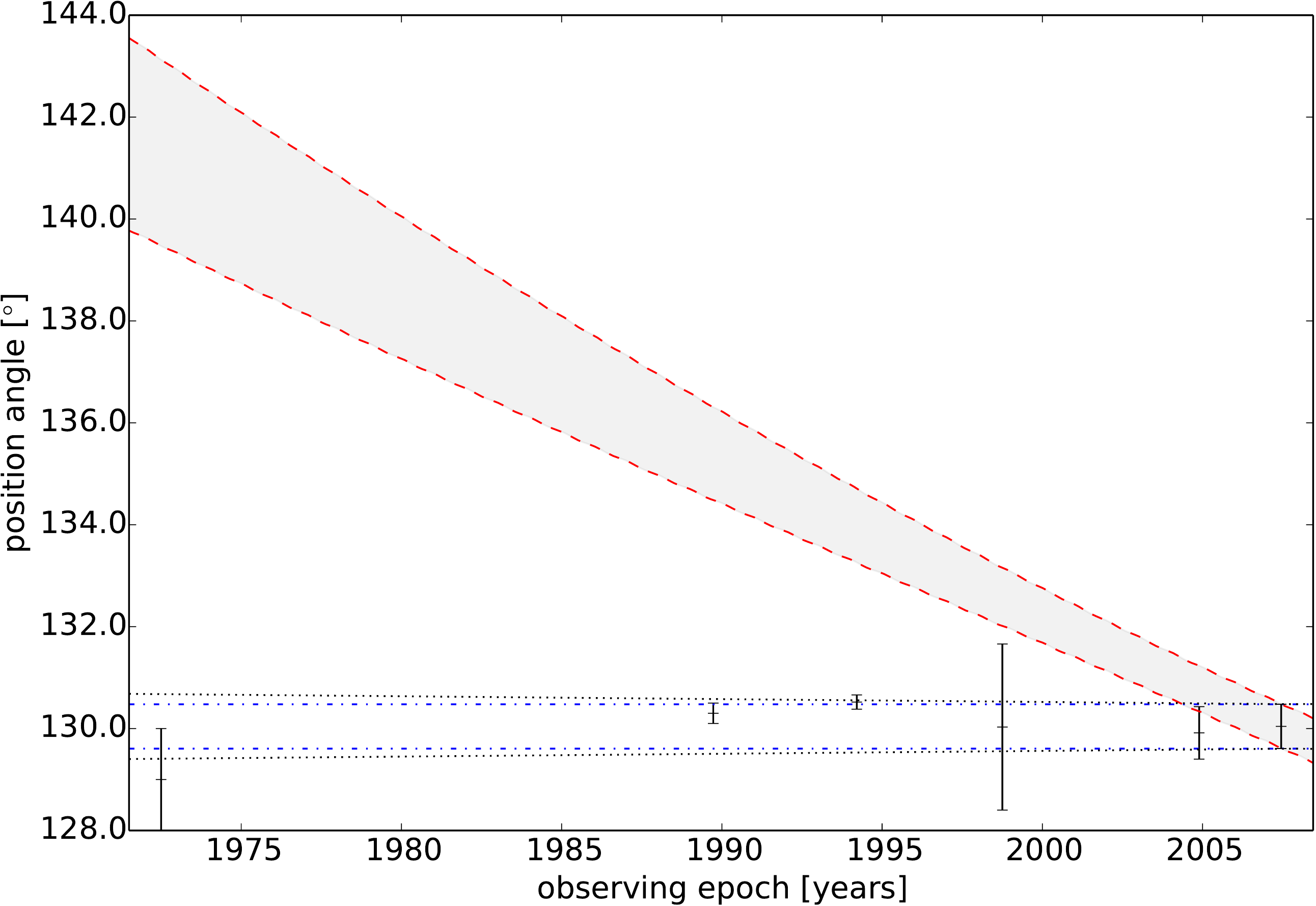}}
\vspace{0.4cm}

\Large{\underline{Proper motion diagram for HIP\,3678\,C}}
\vspace{0.4cm}

\resizebox{\hsize}{!}{\includegraphics{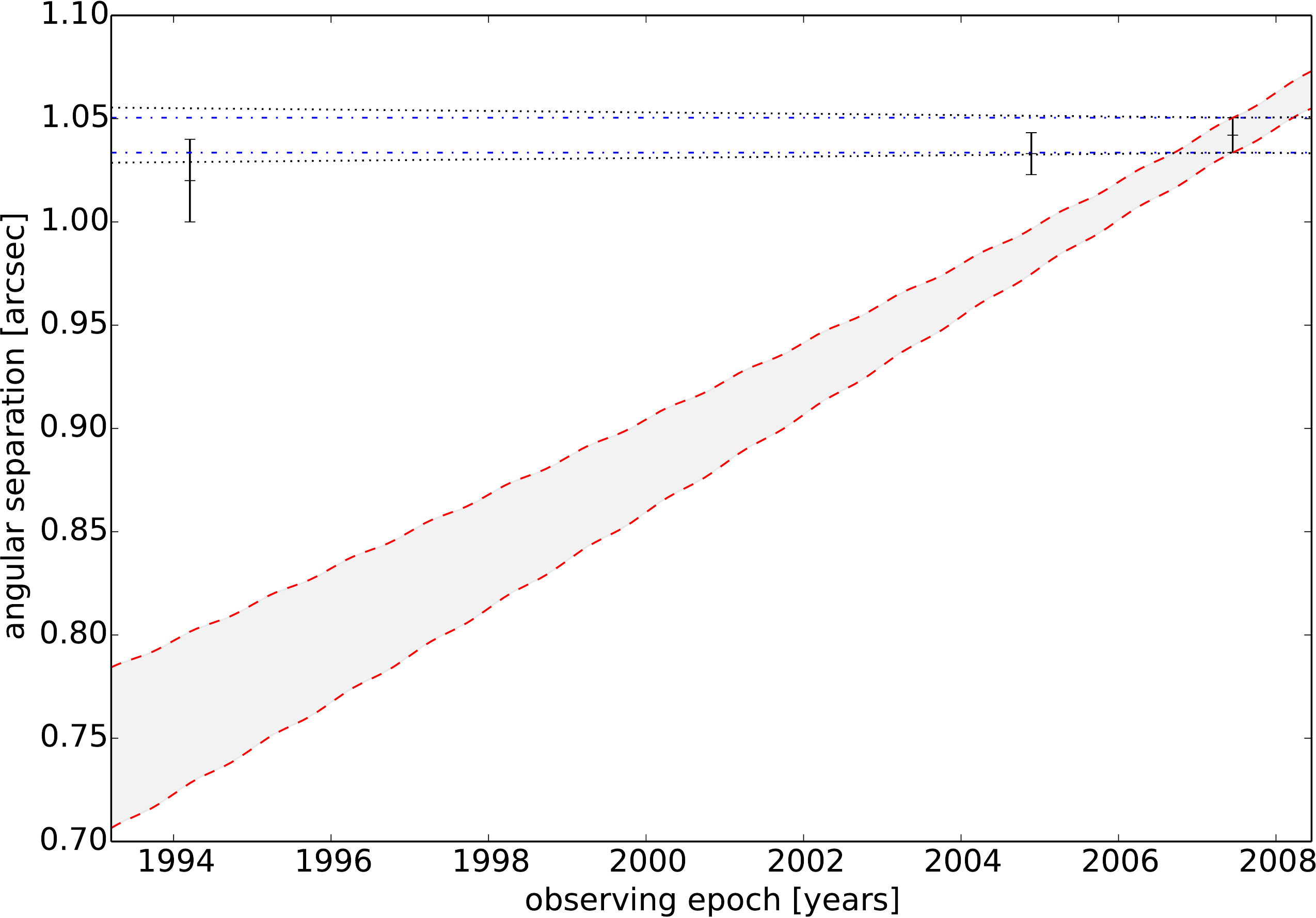}\hspace{10mm}\includegraphics{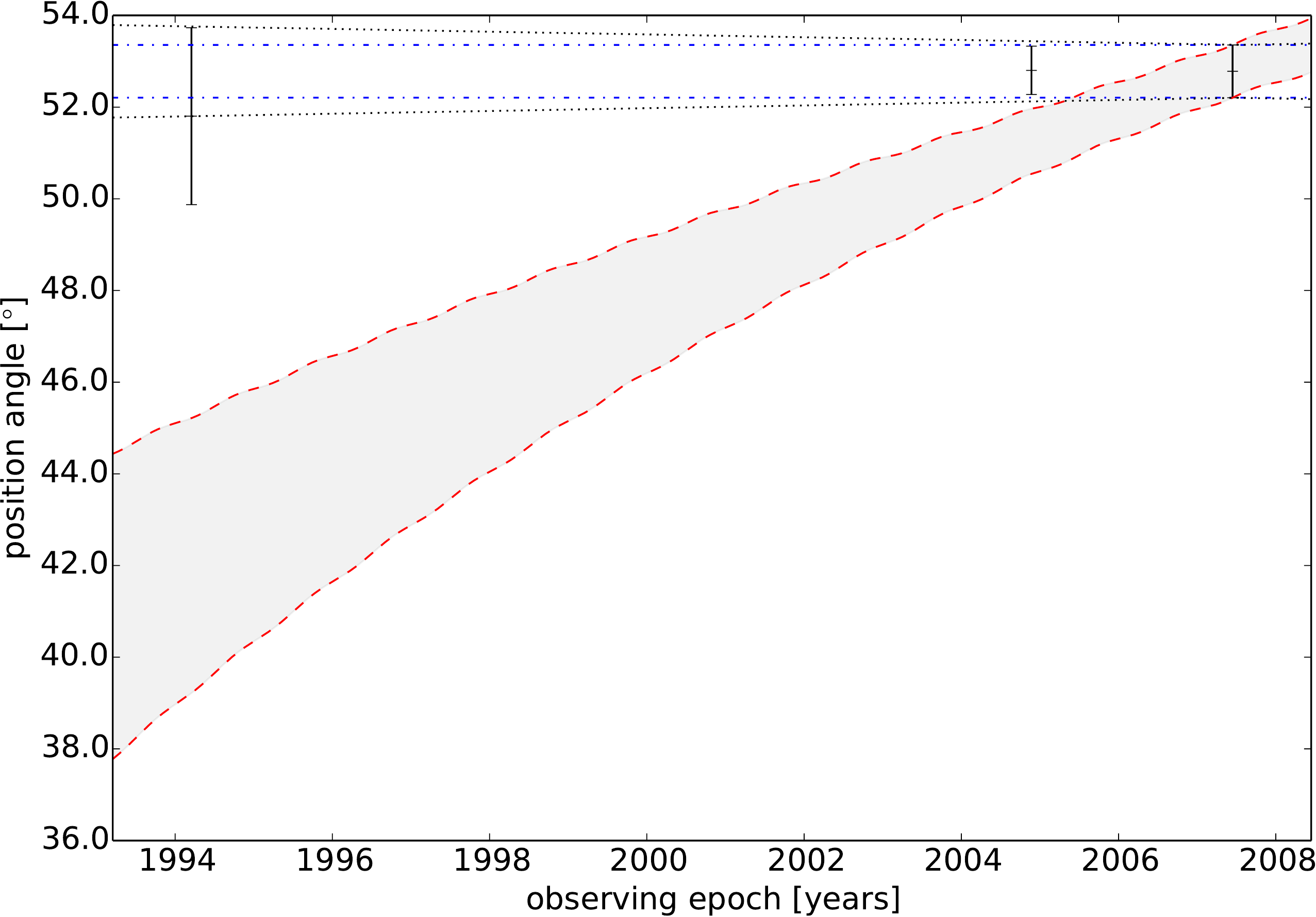}}

\caption{The proper motion diagrams for angular separation (left) and position angle (right) over time for HIP\,3678\,B (top), and the newly detected co-companion HIP\,3678\,C (bottom), respectively. The area of maximal possible orbital motion for a circular edge-on (for separation), or a pole-on (for position angle) orbit of a physically bound system are indicated by dotted (black) lines. The dashed (red) lines mark the (greyish) area where we would expect a non-moving background object, derived with the known proper and parallactic motion of HIP\,3678\,A.}
\label{fig:cpm}
\end{figure*}

By comparing the measured astrometry of HIP\,3678\,B and of the detected companion-candidate with the expected astrometry for non-moving background sources, we can check for common proper motion of both objects. The results of this common proper motion analysis are summarized in Tab.\,\ref{tab:abs_astrometry} and are illustrated in Fig.\,\ref{fig:cpm}.

Beside the \textit{HST} observation taken in 1994 and the NACO imaging data from 2004 and 2007, two further astrometric measurements of HIP\,3678\,B are given in the literature \citep[][]{cudworth1973, bond1999}. In addition, the companion is also detected together with its primary star in the 2 Micron All Sky Survey (2MASS), and the astrometric positions of both objects are listed in the 2MASS Point Source Catalog (2MASS-PSC), see \cite{skrutskie2006}.

HIP\,3678\,B is located about 3.9\,arcsec south-east of HIP\,3678\,A at a position angle of about 130\,$\degr$. Within the given total epoch difference between the first and the latest observing epoch of about 35\,yr, the astrometry of HIP\,3678\,B significantly deviates from the expected one for a non-moving background object. We can reject this background hypothesis for the companion at a significance level of more than 8\,$\sigma$, in total (2.8\,$\sigma$ in angular separation and 5.9\,$\sigma$ in position angle, respectively). Hence, our astrometric analysis significantly confirms that HIP\,3678\,B is a common proper motion companion of the white dwarf HIP\,3678\,A. Assuming a physical relation of both stars, at a distance of 504\,pc the measured angular separation of the comoving companion corresponds to a projected separation of about 1900\,au, which yields a minimal orbital period of about 60000\,yr, assuming a total mass of the HIP\,3678 system of about 1.8\,M$_{\sun}$ (see below).

In contrast to HIP\,3678\,B, the newly found faint companion-candidate is imaged only in three observing epochs with a total epoch difference of about 13\,yr. According to the common proper motion analysis, we can reject the background hypothesis for this candidate at a significance level of more than 8\,$\sigma$ in total (6.1\,$\sigma$ in angular separation and 2.7\,$\sigma$ in position angle, respectively). Hence, this object is a further comoving companion of HIP\,3678\,A, which therefore will be designated as HIP\,3678\,C, from hereon. At the distance of the HIP\,3678 system, the angular separation of HIP\,3678\,C relative to its primary star is about 500\,au (minimal orbital period of about 11000\,yr for an assumed total mass of the HIP\,3678\,AC system of about 1\,M\,$_{\sun}$, see below), which makes the central star of the planetary nebula NGC\,246 a hierarchical triple, composed of a close binary (including the white dwarf), which exhibits a further companion at a wider separation.

For both HIP\,3678\,B and C, we did not find any significant drifts in angular separation and/or position angle, which could be an indication for detected orbital motion, as it is also expected by taking into account the given astrometric uncertainties and long orbital periods of the companions.

\begin{table*}

\caption{Astrometry of the HIP\,3678 system. We show the measured angular separation and position angle of HIP\,3678\,B \& C relative to their primary star HIP\,3678\,A for all observing epochs. In the columns Sig.-level(background), we list the significance level to reject the background hypothesis for both companions relative to the latest observing epoch.}
\label{tab:abs_astrometry}
\begin{tabular}{@{}lccccccc}
\hline\hline
Object        & Observing          & Reference & angular    & Sig.-level & position & Sig.-level\\
              & Epoch              &           & separation & (background)     & angle    & (background)    \\
              & (YYYY-MM-DD)       &           & (arcsec)   & ($\sigma$) & (\degr)  & ($\sigma$)\\

\hline
HIP\,3678\,B  & 1972-07-02 & \citet{cudworth1973}    & $3.80 \pm 0.10$ & $2.8$        & $129.0 \pm 1.0$  & $5.9$\\
              & 1989-09-24 & \citet{bond1999}        & $3.81 \pm 0.01$ & $3.2$        & $130.3 \pm 0.2$  & $5.5$\\
              & 1994-03-17 &                         & $3.84 \pm 0.02$ & $3.2$        & $130.5 \pm 0.2$  & $4.7$\\
              & 1998-10-02 & 2MASS-PSC               & $3.83 \pm 0.12$ & $0.6$        & $130.1 \pm 1.6$  & $1.5$\\
              & 2004-11-22 &                         & $3.85 \pm 0.04$ & $0.4$        & $129.9 \pm 0.5$  & $1.3$\\
              & 2007-06-14 &                         & $3.90 \pm 0.03$ & $-$          & $130.1 \pm 0.4$  & $-$  \\

HIP\,3678\,C  & 1994-03-17 &                         & $1.02 \pm 0.02$ & $6.1$        & $51.8 \pm 1.9$   & $2.7$\\
              & 2004-11-22 &                         & $1.03 \pm 0.10$ & $3.2$        & $52.8 \pm 0.5$   & $1.7$\\
              & 2007-06-14 &                         & $1.05 \pm 0.01$ & $-$          & $52.9 \pm 0.6$   & $-$  \\
\hline
 \end{tabular}
\end{table*}

\section{Photometry}
\label{sec:phot}

Photometric measurements of HIP\,3678\,A and of its two comoving companions were obtained with NACO during the imaging epoch in 2004. Observations were carried out in the \textit{J}, \textit{H}, and \textit{Ks} band and the photometric standard star GSPC\,S677-D\footnote{$\rmn{\alpha}(J2000)=23^{\rmn{h}} 23^{\rmn{m}} 34\fs5$, $\rmn{\delta}(J2000)=-15\degr 21\arcmin 06\arcsec$}, whose photometry is listed in the 2MASS-PSC\footnote{2MASSJ\,23233432-1521094: $J = 11.851 \pm 0.021$\,mag, $H = 11.558 \pm 0.024$\,mag, and $Ks = 11.507 \pm 0.025$}, was observed in the same night at airmass differences to the HIP\,3678 system, which were always smaller than 0.03\,dex.

The instrumental magnitudes of the individual components of the HIP\,3678 system, as well as those of the standard star are measured in all bands with aperture photometry using the {\small IDL}\verb"/aper" routine. In the case of the close binary HIP\,3678\,AC, the contamination of the photometry due to the components on each other is taken into account by subtracting the individual point spread functions of the objects via radial filtering. The absolute magnitudes of the components of the HIP\,3678 system are derived with the measured apparent magnitudes, adopting a distance of $504 \pm 178\,$\,pc. The obtained $J-K$-colours and absolute magnitudes of the components are listed in Tab.\,\ref{tab:phot}. The colours are converted in the CIT photometric system \citep{elias1983}, using the transformations equations from \citet{carpenter2001}.

\begin{table}
\caption{The measured apparent photometry of all components of the HIP\,3678 system. The absolute photometry of the components are derived with the apparent photometry of the objects and with the distance of the central star of the planetary nebula NGC\,246. The $J-K$-colour of all objects is given in the CIT photometric system.}\label{tab:phot}
\begin{tabular}{cccc}
\hline\hline
Object            & \textit{J}               & \textit{H}                & \textit{Ks}              \\
	              & (mag)            & (mag)            & (mag)           \\
\hline
HIP\,3678\,A      & $12.81 \pm 0.02$ & $12.90 \pm 0.02$ & $12.90 \pm 0.03$\\
HIP\,3678\,B      & $13.07 \pm 0.02$ & $12.62 \pm 0.02$ & $12.46 \pm 0.03$\\
HIP\,3678\,C      & $18.44 \pm 0.05$ & $17.94 \pm 0.06$ & $17.53 \pm 0.05$\\
\hline
$J-K$             & $M_{J}$     & $M_{H}$  & $M_{Ks}$\\
$[$mag$]$         & (mag)            & (mag)         & (mag)\\
\hline

$-0.07 \pm 0.04$\,\,\,\,             & $4.3 \pm	0.8$        & $4.4 \pm 0.8$ & $4.4 \pm 0.8$\\
$0.60 \pm 0.04$                      & $4.6 \pm 0.8$        & $4.1 \pm 0.8$ & $3.9 \pm 0.8$\\
$0.87 \pm 0.08$                      & $9.9 \pm 0.8$        & $9.4 \pm 0.8$ & $9.0 \pm 0.8$\\
\hline
\end{tabular}
\end{table}

\begin{figure}
\resizebox{\hsize}{!}{\includegraphics{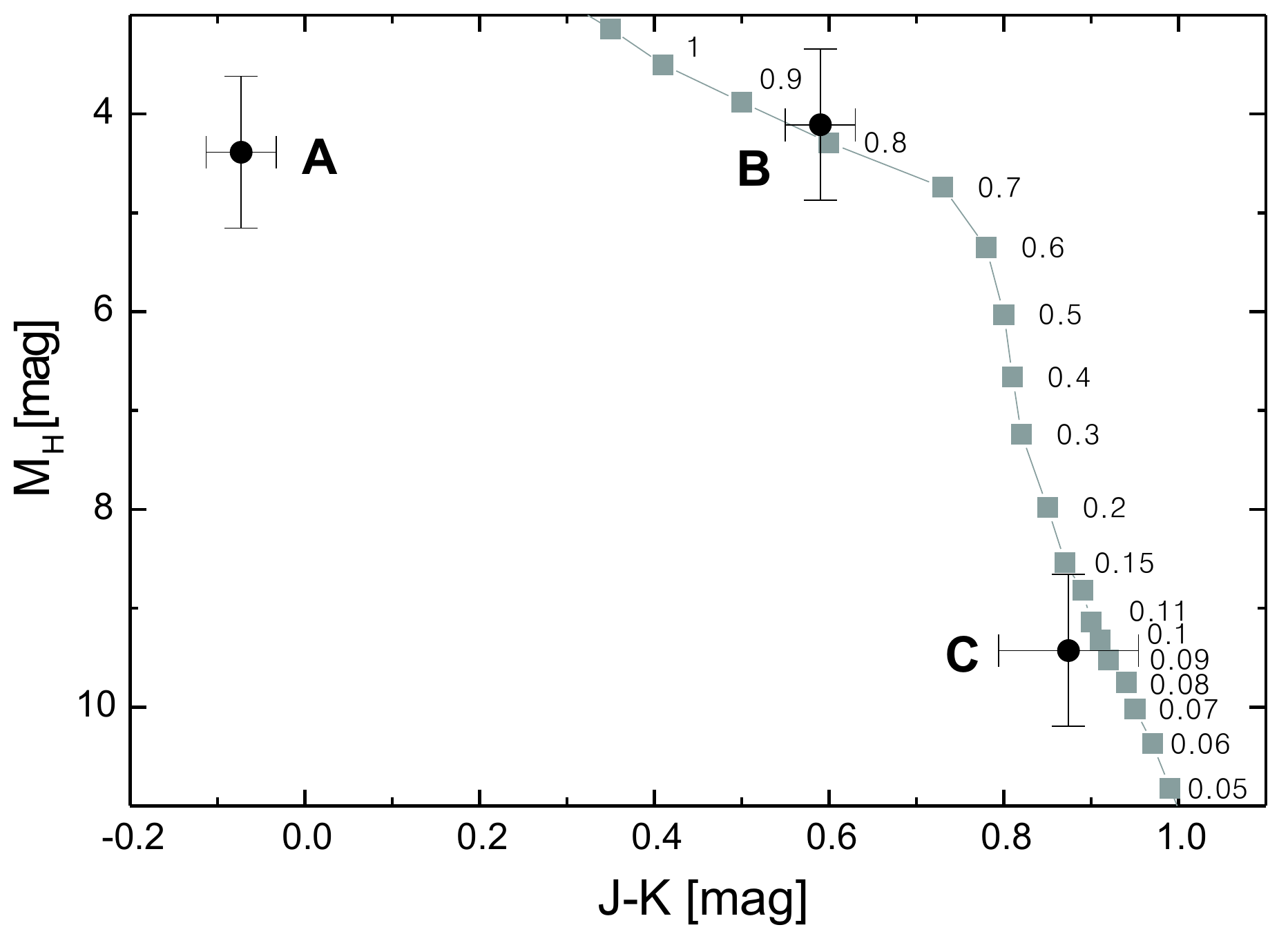}}
\caption{Colour-magnitude diagram for all members of the HIP\,3678 system. Shown is the absolute \textit{H} band magnitude versus the $J-K$-colour of all objects in the CIT system. In addition, we plot the isochrone for $\log(\rm{age[yr]}) = 8.4$ of the evolutionary models from \citet{baraffe1998}. The photometry of HIP\,3678\,B \& C agrees well with two low-mass stars with an age of 260\,Myr, which are located at the distance of the HIP\,3678 system.}
\label{img:CMD}
\end{figure}

With a mass of 0.84\,M$_{\sun}$ for the white dwarf HIP\,3678\,A, as derived by \cite{koesterke1998}, we expect a mass of its progenitor star of about 4.3$\,\rmn{M_{\sun}}$, using the initial to final mass relationship of white dwarf from \citet{catalan2008}. This yields a main-sequence lifetime of the progenitor star of about 260\,Myr (assuming a mass--luminosity relation of $L \propto M^{3.5}$). Due to the short cooling age of the white dwarf of only about 6600\,yr, as derived by \cite{ali2012}, the derived age of the progenitor star also corresponds to the total age of the HIP\,3678 system, assuming that the HIP\,3678\,A, B \& C are coeval. In the colour--magnitude diagram in Fig.\,\ref{img:CMD}, we show the colours and absolute magnitudes of all components of the HIP\,3678 system together with the $\log(\rm{age[yr]})=8.4$ isochrone of the evolutionary models of low-mass star from \citet{baraffe1998}.

While HIP\,3678\,A appears significantly bluer than a main-sequence star of the same absolute magnitude, as it is expected for a young and hot white dwarf, the photometry of the two comoving companions HIP\,3678\,B and C agrees well with the expected colours and magnitudes of low-mass stars with an age of 260\,Myr, which are located at the distance of the HIP\,3678 system. Hence, the companionship of both comoving companions to HIP\,3678\,A is well supported by photometry.

By assuming a system age of 260\,Myr, we can derive the masses of HIP\,3678\,B and C using the determined absolute magnitudes of the comoving companions and the \cite{baraffe1998} evolutionary models, which yield a mass of $0.85 \pm 0.11$\,M$_{\sun}$ for HIP\,3678\,B and $0.098 \pm 0.024$\,M$_{\sun}$ for HIP\,3678\,C, respectively. Furthermore, the spectral types of the detected companions can be approximated using the magnitude--spectral-type relation from \citet{reid2004}. According to this relation, we expect that HIP\,3678\,B is an early to mid-K dwarf (K2--K5), while HIP\,3678\,C is a mid-M dwarf with a spectral type in the range between M5 and M6.

\section{Detection Limits and further companions of the HIP\,3678 system}

\begin{figure}
\resizebox{\hsize}{!}{\includegraphics{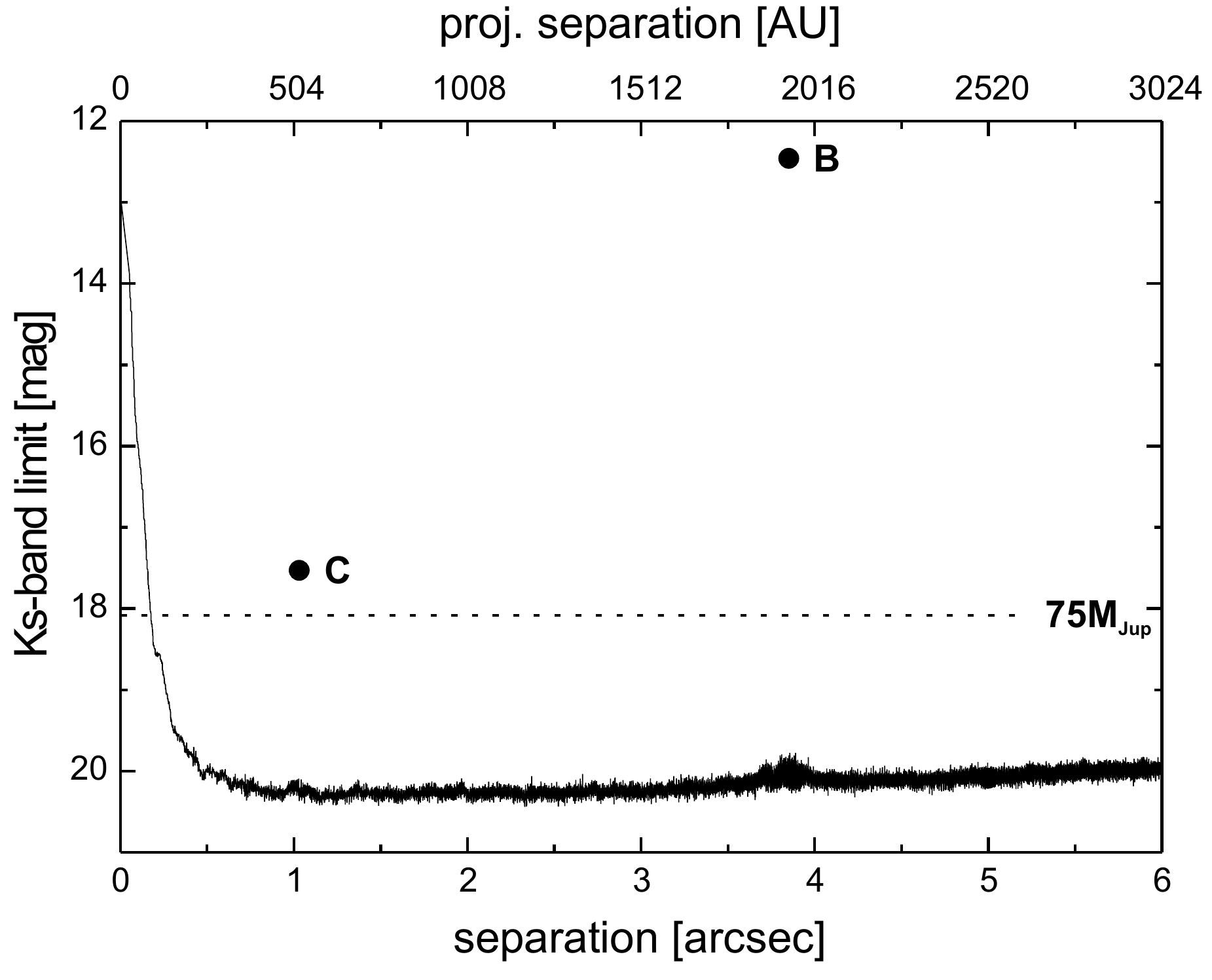}}
\caption{The achieved ($S/N=3$) detection limit of the NACO \textit{Ks} band observation taken in 2004, plotted versus angular (below) and projected (top) separation to HIP\,3678\,A. A limiting magnitude of about 20\,mag is reached in the background-limited region beyond 0.5\,arcsec, where brown dwarf companions with masses down to 36\,M$_{\rm{Jup}}$ can be detected with projected separations to HIP\,3678\,A of more than about 250\,au. Beside the two comoving companion HIP\,3678\,B and C, additional stellar companions of HIP\,3678\,A can be excluded with projected separation larger than 130\,au up to the field of view, fully covered by the NACO observation (angular separations of up to 6\,arcsec or about 3000\,au of projected separation).}\label{img:DetLim}
\end{figure}

Among all imaging data presented in this work, the NACO \textit{Ks} band observations from 2004 exhibit the highest contrast. The achieved ($S/N=3$) detection limit versus angular and projected separation is illustrated in Fig.\,\ref{img:DetLim}.

Beside HIP\,3678\,B \& C, no additional companion-candidates could be detected in this observing epoch within the fully covered field of view, i.e. at angular separations smaller than 6\,arcsec ($\sim$\,3000\,au of projected separation) around HIP\,3678\,A. In the background noise limited region a ($S/N=3$) detection limit\footnote{Choice of $S/N=3$, confirmed by inserting and retrieving simulated companions at this contrast level, see \citep{haase09}.} of about 20\,mag is reached at angular separations from HIP\,3678\,A beyond about 0.5\,arcsec. At the derived age of the HIP\,3678 system of about 260\,Myr, this allows the detection of low-mass sub-stellar companions with masses down to 36\,$M_{\rm{Jup}}$ and projected separations from HIP\,3678\,A of more than about 250\,au. All stellar companions (mass$ > 75$\,M$_{\rm{Jup}}$) of HIP\,3678\,A can be detected at angular separations larger than 0.26\,arcsec ($\sim$130\,au of projected separation). The NACO observations from 2007 were taken with the S27 optics and fully cover a field of view around HIP\,3678\,A with an angular radius of 11\,arcsec ($\sim$\,5500\,au of projected separation). Also in this larger field of view, no additional companions of HIP\,3678\,A are detected. Due to worse weather conditions, these imaging data are about 0.5\,mag less sensitive than the ones taken in 2004, but the achieved contrast is sufficiently high so that additional stellar companions can be ruled out at angular separations beyond 0.45\,arcsec ($\sim$\,220\,au) and companions with masses down to 43\,M$_{\rm{Jup}}$ are detectable in the background-limited region at angular separations wider than 1\,arcsec ($\sim$\,500\,au). Combined with the high-contrast imaging data, obtained with NACO in 2004, we therefore can conclude that there are no additional stellar companions of HIP\,3678\,A with projected separations between 130 and up to 5500\,au.

\section{Results and Discussion}
\label{sec:sum}

In the course of our multiplicity study of B stars, we have analysed near-infrared and visual imaging data taken with NACO/VLT and WFPC2/\textit{HST} of HIP\,3678\,A the central star of the planetary nebula NGC\,246. The individual data were taken from the ESO and \textit{HST} data-archives, and were combined with data points from the literature and from the 2MASS-PSC. In the high-contrast NACO, as well as in the HST data of lower resolution, we detected a new companion of HIP\,3678\,A, which clearly shares a common proper motion with its primary star. The new companion HIP\,3678\,C is located north-east of HIP\,3678\,A at an angular separation of about 1\,arcsec ($\sim$\,500\,au of projected separation). With its previously known comoving companion, this detection makes the central star of the planetary nebula an hierarchical triple system, composed of the close binary system HIP\,3678\,AC and its wider companion HIP\,3678\,B at a projected separation of about 1900\,au ($\sim$\,3.9\,arcsec). As described in the literature HIP\,3678\,A exhibits a mass of about 0.84\,M$_{\sun}$, as derived from its effective temperature and luminosity using evolutionary models of young white dwarfs. With the initial to final mass relation of white dwarfs from \cite{catalan2008}, this yields an initial mass of the progenitor star of HIP\,3678\,A of about 4.3$\,\rmn{M_{\sun}}$. Due to the short dynamical age of the planetary nebula NGC\,246 of about only 6600\,yr, as described in the literature, the age of the HIP\,3678 system can be approximated to correspond to the main-sequence lifetime of the white dwarf progenitor star of about 260\,Myr. The obtained apparent photometry of the comoving companions HIP\,3678B \& C agrees well with low-mass stellar companions with the same age, as derived for HIP\,3678\,A, being located at the distance of the white dwarf. According to the \cite{baraffe1998}, evolutionary models HIP\,3678\,B and C exhibit masses of about 0.85 and 0.1\,M$_{\sun}$, respectively. Using the absolute magnitude--spectral-type relation from \cite{reid2004}, we expect that HIP\,3678\,B is an early to mid-K dwarf, while HIP\,3678\,C is a M5 to M6 dwarf.

With the NACO high-contrast imaging observations, additional stellar companions of the white dwarf can be ruled out around HIP\,3678\,A at projected separations in the range between 130 and up to 5500\,au. The same holds for brown dwarf companions with masses down to about 36\,M$_{Jup}$ at projected separations beyond about 500 up to 3000\,au, and down to 36\,M$_{Jup}$ in the range of separation between 3000 and up to 5000\,au.

The HIP\,3678 is an evolved stellar system, which underwent a significant mass-loss of its primary component. Assuming conservation of angular momentum e.g. due to radial symmetric mass-loss during the post main-sequence lifetime of the white dwarf progenitor star and the formation phase of the planetary nebula, we can approximate the initial separations of the components to HIP\,3678\,A. The total mass of the system decrease from about 5.2 to 1.8\,M$_{\sun}$, which yields an expansion factor of the HIP\,3678 system of about three. Hence, the system was significantly smaller in its initial configuration, but clearly wider than about 100\,au. Due to the wide initial separation of the HIP\,3678\,AC binary, a common envelope phase of the evolved white dwarf progenitor star and its M dwarf companion during the asymptotic giant branch phase of the star can most probably be ruled out.

So far, about 40 close binaries could be detected as central stars of planetary nebulae all, which exhibit orbital periods of up to only a few days, mostly detected by periodic photometric variability or excess emission in the near-infrared \citep[see][ De Marco 2014, and references therein]{demarco2013}\nocite{demarco2014}. Wider binaries with periods of more than 1000\,d could be detected in the centre of planetary nebulae via radial velocity measurements \citep[see e.g.][]{vanwinckel2014}.

In addition to these close stellar systems, only a few planetary nebulae are known, which harbour binary systems in their centres with wider separations of a few hundreds up to many thousands au, i.e. with orbital periods of more than 1000\,yr. The main survey for such wide systems was carried out with the \textit{HST} \citep{ciardullo1999}. More recently further wide binaries in the centre of planetary nebulae could be identified, e.g. by \citet{benetti2003}, or most recently by \citet{liebert2013}. Among these nebulae, there are also two with possible triples in their centre, namely Abell\,63 and NGC\,7008. However, the triple nature of the central stars of these nebulae needs confirmation via follow-up high-contrast imaging observations and astrometry. Hence, the detection of HIP\,3678\,C, reported here, makes NGC\,246 the first known planetary nebula with a confirmed hierarchical triple stellar system in its centre.

\section*{Acknowledgements}
CA would like to thank the German Science Foundation (DFG) \& SPP\,1385 for support in grant NE 515/35-1. MM wants to thank DFG for support in grant MU 2695/13-1. This research has made use of the SIMBAD and VizieR data bases, operated by at CDS in Strasbourg, France. Based on observations made with the NASA/ESA \textit{HST}, and obtained from the Hubble Legacy Archive, which is a collaboration between the Space Telescope Science Institute (STScI/NASA), the Space Telescope European Coordinating Facility (ST-ECF/ESA) and the Canadian Astronomy Data Center (CADC/NRC/CSA). Based on photographic data obtained using The UK Schmidt Telescope. The UK Schmidt Telescope was operated by the Royal Observatory Edinburgh, with funding from the UK Science and Engineering Research Council, until 1988 June, and thereafter by the Anglo-Australian Observatory. Original plate material is copyright (c) the Royal Observatory Edinburgh and the Anglo-Australian Observatory. The plates were processed into the present compressed digital form with their permission. The Digitized Sky Survey was produced at the Space Telescope Science Institute under US Government grant NAG W-2166.

\end{document}